\documentclass[fleqn,12pt,twoside]{article}
\usepackage{espcrc1}

\usepackage{graphicx}
\usepackage{epsfig}
\usepackage[figuresright]{rotating}

\newcommand{\ttbs}{\char'134}
\newcommand{\AmS}{{\protect\the\textfont2
  A\kern-.1667em\lower.5ex\hbox{M}\kern-.125emS}}

\begin{document}

\def\lsim{{\buildrel < \over\sim}}
\def\gsim{{\buildrel > \over\sim}}
\def\to{\rightarrow}
\def\fb{~{\rm fb}}
\def\pb{~{\rm pb}}
\def\ev{\,{\rm eV}}
\def\mev{\,{\rm MeV}}
\def\gev{\,{\rm GeV}}
\def\tev{\,{\rm TeV}}
\def\wh{\widehat}
\def\wt{\widetilde}
\def\mhalf{m_{1/2}}
\def\gl{\wt g}
\def\q{$q$}
\def\qbar{$\bar{q}$}
\def\g{$g$}
\def\dc{$\delta_c$}
\def\als{\alpha_s}
\newcommand{\as}{{\ifmmode \alpha_S \else $\alpha_S$ \fi}}
\def\ttbs{\char'134}
\def\AmS{{\protect\the\textfont2
  A\kern-.1667em\lower.5ex\hbox{M}\kern-.125emS}}
%%%%%%%%%%%%%%%%%%%%%%%%%%%%%%%%%%%%%%%%%%%%%%%%
%%%%%%%%%%%%%%%%%%%%%%%%%%%%%%%%%%%%%
%%%%%%%%%%%%%%%%%%%%%%%%%%%%%%%%%%%%%%%%
\catcode`@=11
\def\Biggg#1{\hbox{$\left#1\vbox to 22.5\p@{}\right.\n@space$}}
\catcode`@=12
\newcommand\refq[1]{$^{#1}$}
\newcommand\ind[1]{_{\rm #1}}
\newcommand\aopi{\frac{\as}{\pi}}
\newcommand\oot{\frac{1}{2}}
\newcommand\sinsthw{\sin^2(\theta_{\rm W})}
\newcommand\logmu{\log(\mu^2/\Lambda^2)}
\newcommand\Lfb{\Lambda^{(5)}}
\newcommand\Lfc{\Lambda^{(4)}}
\newcommand\Lf{\Lambda_5}
\newcommand\epem{\ifmmode e^+e^- \else $e^+e^-$ \fi}
\newcommand\mupmum{ \mu^+\mu^- }
\newcommand\ms{\ifmmode{\overline{\rm MS}}\else $\overline{\rm MS}$\ \fi}
\newcommand\Q[1]{_{\rm #1}}
\newcommand\pplus[1]{\left[\frac{1}{#1}\right]_+}
\newcommand\plog[1]{\left[\frac{\log(#1)}{#1}\right]_+}
\newcommand\sh{\hat{s}}
\newcommand\epbar{\overline\epsilon}
\newcommand\nf{\alwaysmath{{n_{\rm f}}}}
\newcommand\MSB{\ifmmode{\overline{\rm MS}}\else $\overline{\rm MS}$\ \fi}
\newcommand{\aem}{\alpha_{\rm em}}
\newcommand{\nlf}{\alwaysmath{{n_{\rm lf}}}}
\newcommand{\ep}{\epsilon}
\newcommand{\aop}{\frac{\as}{2 \pi}}
\newcommand{\Tf}{{T_{\rm f}}}
\newcommand{\mub}{\ifmmode \mu{\rm b} \else $\mu{\rm b}$ \fi}
\newcommand\alwaysmath[1]{\ifmmode #1 \else $#1$ \fi}
\newcommand{\TeV}{{\rm TeV}}
\newcommand{\GeV}{{\rm GeV}}
\newcommand{\MeV}{{\rm MeV}}
\newcommand{\LQCD}{\ifmmode \Lambda_{\rm QCD} \else $\Lambda_{\rm QCD}$ \fi}
\newcommand{\LMSB}{\ifmmode \Lambda_{\overline{\rm MS}} \else
          $\Lambda_{\overline{\rm MS}}$ \fi}
\newcommand{\qb}{\overline{q}}
\def\pp{\ifmmode p\bar{p} \else $p\bar{p}$ \fi}
\def\VEV#1{\left\langle #1\right\rangle}
\def\LMSb{\ifmmode \Lambda_{\rm \overline{MS}} \else
$\Lambda_{\rm \overline{MS}}$ \fi}
%       This defines et al., i.e., e.g., cf., etc.
\def\ie{\hbox{\it i.e.}{}}      \def\etc{\hbox{\it etc.}{}}
\def\eg{\hbox{\it e.g.}{}}      \def\cf{\hbox{\it cf.}{}}
\def\etal{\hbox{\it et al.}}
\def\dash{\hbox{---}}
\def\abs#1{\left| #1\right|}

\def\to{\rightarrow}
%%%%%%%%%%%%%%%%%%
\newcommand{\lra}{\leftrightarrow}
\newcommand{\la}{\langle}
\newcommand{\dd}{{\rm d}}
\newcommand{\PS}{{\rm PS}}
\newcommand{\pperp}{p_{\perp}}
\newcommand{\ra}{\rangle}
\def\vspaceinarray{\nonumber ~&~&~\\}
%%%%%%%%%%%%%%%%%%%%%%%%%%%%%%

\newcommand{\Nc}{N_c}
\newcommand{\Nf}{N_f}
\def\LO{leading order }
\def\CDR{conventional dimensional regularization }
\def\DR{dimensional reduction }
\def\A#1#2{\la#1#2\ra}
\def\B#1#2{[#1#2]}
\def\s#1#2{s_{#1#2}}
\def\h#1#2#3#4{\la#1#2#3#4\ra}
\def\f#1#2#3#4{(#1#2#3)_#4}
\def\P#1#2{{\cal P}_{#1#2}}
\def\L#1#2{\left\{#2\right\}_{#1}}
\def\C#1#2#3#4{\left\{^{#1#2}_{#3#4}\right\}}
\newcommand{\Lmin}{{\rm L}}
\def\Li{{\rm Li_2}}
\def\del#1{\lower.25em\hbox{\LARGE $\times$}\kern -1em #1 }
\def\st#1{\lower.25em\hbox{$|_{#1}$} }  
\newcommand{\eps}{\epsilon}
\newcommand{\ve}{\varepsilon}
\newcommand\epb{\overline{\epsilon}}
\newcommand{\beq}{\begin{equation}}
\newcommand{\eeq}{\end{equation}}
\newcommand{\be}{\begin{equation}}
\newcommand{\ee}{\end{equation}}
\newcommand{\beqn}{\begin{eqnarray}}
\newcommand{\eeqn}{\end{eqnarray}}
\newcommand{\bea}{\begin{eqnarray}}
\newcommand{\eea}{\end{eqnarray}}
\newcommand{\beqns}{\begin{eqnarray*}}
\newcommand{\eeqns}{\end{eqnarray*}}
\def\abs#1{\left| #1\right|}
\def\Am{{\cal A}}
\def\nn{\nonumber}
\def\phys{{\rm phys}}
\def\ms{$\overline{{\rm MS}}$}
\def\limes#1{\mathrel{\mathop{\lim}\limits_{#1}}}
\def\arrowlimit#1{\mathrel{\mathop{\longrightarrow}\limits_{#1}}}
\def\mus#1#2{\left(-\frac{\mu^2}{s_{#1#2}}\right)^\varepsilon}
\def\qb{\bar{q}}
\def\Qb{\bar{Q}}
\newcommand{\y}{\gamma}
\newcommand{\yf}{\gamma_5}
\newcommand{\yh}{\hat{\gamma}}
\newcommand{\yt}{\tilde{\gamma}}
\newcommand{\yb}{\bar{\gamma}}
\newcommand{\cg}{c_\Gamma}

%       This defines et al., i.e., e.g., cf., etc.
\def\ie{\hbox{\it i.e.}{}}      \def\etc{\hbox{\it etc.}{}}
\def\eg{\hbox{\it e.g.}{}}      \def\cf{\hbox{\it cf.}{}}
\def\etal{\hbox{\it et al.}}    \def\vs{\hbox{\it vs.}{}}
\def\dash{\hbox{---}}

\relax
\def\ap#1#2#3{
        {\it Ann. Phys. (NY) }{\bf #1} (19#3) #2}
\def\app#1#2#3{
        {\it Acta Phys. Pol. }{\bf #1} (19#3) #2}
\def\ar#1#2#3{
        {\it Ann. Rev. Nucl. Part. Sci. }{\bf #1} (19#3) #2}
\def\cmp#1#2#3{
        {\it Commun. Math. Phys. }{\bf #1} (19#3) #2}
\def\cpc#1#2#3{
        {\it Comput. Phys. Commun. }{\bf #1} (19#3) #2}
\def\ijmp#1#2#3{
        {\it Int .J. Mod. Phys. }{\bf #1} (19#3) #2}
\def\ibid#1#2#3{
        {\it ibid }{\bf #1} (19#3) #2}
\def\jmp#1#2#3{
        {\it J. Math. Phys. }{\bf #1} (19#3) #2}
\def\jetp#1#2#3{
        {\it JETP Sov. Phys. }{\bf #1} (19#3) #2}
\def\ib#1#2#3{
        {\it ibid. }{\bf #1} (19#3) #2}
\def\mpl#1#2#3{
        {\it Mod. Phys. Lett. }{\bf #1} (19#3) #2}
\def\nat#1#2#3{
        {\it Nature (London) }{\bf #1} (19#3) #2}
\def\np#1#2#3{
        {\it Nucl. Phys. }{\bf #1} (19#3) #2}
\def\npsup#1#2#3{
        {\it Nucl. Phys. Proc. Sup. }{\bf #1} (19#3) #2}
\def\pl#1#2#3{
        {\it Phys. Lett. }{\bf #1} (19#3) #2}
\def\pr#1#2#3{
        {\it Phys. Rev. }{\bf #1} (19#3) #2}
\def\prep#1#2#3{
        {\it Phys. Rep. }{\bf #1} (19#3) #2}
\def\prl#1#2#3{
        {\it Phys. Rev. Lett. }{\bf #1} (19#3) #2}
\def\physica#1#2#3{
        { Physica }{\bf #1} (19#3) #2}
\def\rmp#1#2#3{
        {\it Rev. Mod. Phys. }{\bf #1} (19#3) #2}
\def\sj#1#2#3{
        {\it Sov. J. Nucl. Phys. }{\bf #1} (19#3) #2}
\def\zp#1#2#3{
        {\it Zeit. Phys. }{\bf #1} (19#3) #2}
\def\tmf#1#2#3{
        {\it Theor. Math. Phys. }{\bf #1} (19#3) #2}

% declarations for front matter
\title{Jet physics at hadron colliders}

\author{Zoltan Kunszt \address{Institute for Theoretical Physics, ETH, Zurich, 
         \\ 
        CH-8093 Zurich, Switzerland}
\thanks{Talk given at "Quark Matter 2002", Nantes, France, July 18-24, 2002}
}

% typeset front matter
\maketitle

\begin{abstract}
I give a short summary of 
jet definition
algorithms and  recent progress in the quantitative 
description of jet production.
\end{abstract}

\section{Introduction}
Over the last decade detailed  theoretical description of 
jet physics   at high energy colliders  has been established    
at the  next-to-leading order (NLO) accuracy
\cite{ERT,EKS,GGK}. The theoretical
predictions   
have been 
 successfully compared with the  experimental data
in the case of three jet production in $\epem$ annihilation, 
one or two jet production in photo-production,
 deep inelastic scattering at HERA, and one or two jet production
at proton-antiproton
collisions at Tevatron~\cite{ESW}.

Jets are footprints of quarks and gluons  produced at
short distances. They  are defined as some collections  of  final state
hadrons such that the  relative angular 
distances between the hadrons in momentum space are small.
 Important cancellation theorems valid in all orders
of perturbation theory suggest that infrared
safe global jet observables
 can quantitatively be described  in terms the 
weak perturbative  dynamics of the point like  partons of QCD.
A jet observable is infrared safe if it does not
 change by adding or removing a soft particle
from the jet or by  splitting 
 an ultra relativistic particle into two collinear
particles  within the jet.

The data with highest jet energy at Tevatron could give us the most stringent
test of the QCD dynamics at short distances and allow for an efficient
search for effects of deviations from 
the Standard Model predictions~\cite{huston-vanc}. 
The data of  CDF and D0 collaborations at Tevatron have 
 about the same ($< 20\% $) or better statistical accuracy as 
the theoretical predictions. At the highest jet energies,  
  the  uncertainties     
in  the jet  energy calibration and the errors of the fitted values
of parton number densities, 
however, lead up to 
systematic errors of  $\approx 50\%$. 

 In the case of three jet production in $\epem$, the experimental
systematic errors are smaller and 
the theoretical uncertainties are considerably larger than the current 
experimental errors~\cite{bethke}. Therefore the theorists have strong
motivations to further improve the accuracy of their predicitons.

Recently, we could witness important theoretical progress  
 in four areas of jet physics.
First, the NLO calculations could be  extended to 5-leg processes
(four jet production in \epem annihilation~\cite{BDK4j,ee4jmcs},
 three jet production 
in hadron-hadron collisions~\cite{nagy3j,BDK5g} \etc).
 Secondly,  remarkable
progress has been made  towards the ambitious goal
 of calculating  jet cross sections of 4-leg
processes  in next-to-next-to leading order
 (NNLO) accuracy~\cite{gehrmann-radcor}.
 In particular, the NNLO virtual corrections
could be calculated analytically  for four leg amplitudes.
Thirdly, the NLO calculations could be improved in the threshold
region by resumming the 
large logarithmic contributions to all order
\cite{sterman-resum}. Finally new techniques
have been developed to calculate many jet processes
in leading order\cite{mangano-moretti}.
Below I discuss difficulties related to  
jet definitions, I briefly review the general theoretical
framework of the NLO calculations, 
 I describe in  some detail  the 
new NLO results obtained for 5-leg processes, finally I summarize the
main results on NNLO corrections.

\section{Jet definition algorithms}

Jet definitions are based on a selection and a recombination 
algorithm. The selection algorithm selects the group of particles which form
the jet and the recombination algorithm specifies how to construct
the kinematicl variables of the jet in terms of the momenta of the 
selected hadrons.

The jet algorithms used in the data analysis
 are more complicated
than the simple theoretical definitions in terms of one, two or three partons.
Some 'auxiliary' features like the phenomena of merging and splitting, or the 
use
of seeds are not always completely specified. These issues may influence
the infrared sensitivity and the size of the hadronic fragmention corrections.
 In a careful  analysis one has to  require 
 that i)~the  jet selection process, 
 the jet kinematic variables, preclustering,
merging, splitting, the role of the underlying evens are fully specified; 
ii)~The fully specifed algorithm should be infrared safe, independent from
the detector properties and the algorithms should be simple to use; 
iii)~The algorithm
has to be defined 
in the same way at parton, hadron and detector level
(for a recent detailed discussion for hadron colliders 
see \cite{betterjetalgo,}).

Although the algorithms have a large amount of arbitrariness 
they should fulfil a number of important theoretical requirements, such as
 insensitivity to soft radiation and to collinear splitting, 
or invariance under boost in the case of hadron colliders \etc .
In the latter case   the use of  transverse momentum and
rapidities as kinematic variables is preferred.
 Another  constraint
is suggested by the study of  resummation corrections. 
Resummations can only be carried out if the  boundaries of the inclusive jet 
kinematic variables  are defined independently
from the number of final state particles~\cite{sterman-resum,CS}.
 This prefers Lorenz covariant
recombination schemes (such as the E-scheme).
 Finally, the algorithm 
 should be as simple as possible. 
At hadron colliders we do not have one best algorithm.
The cone algorithm is broadly used while the $k_T$ algorithm 
is preferred more by the theorists.

\subsection{Cone jet algorithm for hadron-hadron collisions}
The cylindrical shape of the  detectors
 and boost invariance suggest that the kinematics has
to be described in terms
rapidities and azimuthal angles~\cite{EKS,GGK}. 
 In the 2-dimensional $\eta \times \phi$
lego-plot the hadrons or partons constituting the jets lie within
a cone of radius R. The trial cone in the lego plot is given by 
its radius  and the value of
its center  $(\eta^{\rm C}\,,\phi^{\rm C})$.
The cone is adjusted in such a way that
the geometric center of the cone agrees with the $E_T$-weighted
recombined values of the particles within the cone. 
All  particles within the trial cone fulfil 
\beq
  \label{cone}
  \sqrt{(\eta^{\rm C}\, -\, \eta_i)^2 + (\phi^{\rm C}\, -\, 
\phi_i^2)} \leq R\,.
\eeq
A stable cone (protojet) is obtained  if the `` physical'' center of 
the cone  defined
by the recombination algorithm  
  \beq
  \label{conevalues}
 \eta^{\rm R}_{\rm T}
=\frac{\sum_i\eta^{i}_{\rm T} E^{\rm i}_{\rm T}}
{ E^{\rm R}_{\rm T}}
\,,\quad\quad
 \phi^{\rm R}_{\rm T}
=\frac{\sum_i\phi^{i}_{\rm T} E^{\rm i}_{\rm T}}
{ E^{\rm R}_{\rm T}}\,,\quad
 \quad
 E^{\rm R}_{\rm T}=\sum_i E^{i}_{\rm T}
\eeq
coincides with its  geometrical center
 $(\eta^{\rm C}\,,\phi^{\rm C})= (\eta^{\rm R}\,,\phi^{\rm R})$.
In this case it is natural to identify the jet variables with 
the recombined values of the stable cone
 $(E^J_T,\eta^{\rm J}\,,\phi^{\rm J}) 
= (E^{\rm R}_T,\eta^{\rm R}\,,\phi^{\rm R})$.
After identification of the jet as  group of particles
within the stable cone one can construct the jet kinematical variables
with using  some  recombination scheme,  for example
\bea
&& E^J_{x,y,z}\,=\,\sum_i E^{i}_{x,y,z}\\
&&\theta^J\,=\,\arctan \frac{E^J_T}{E^J_z}, \quad \quad
 \phi^J\,=\,\arctan \frac{E^J_x}{E^J_y}, \\
&&\eta^J\,=\,-\ln \left(\tan \frac{\eta^J}{2}\right), 
\quad\quad
 E^J_T\,=\, E^J \sin \theta^J
\eea
Different recombination algorithms give  aproximately the same
kinematical values if $M^J<<E^J_T$. The boost invariant recombination
scheme, however, is a better estimator of the jet variables.
I have noted above that  boost invariant variables are not suitable for 
resummation studies since the kinematic boundary of the $E^J_T$ depends
on the number of partons. Resummation is  consistent with   e.g. 
the  E-scheme which
 recombines the jet four momenta as
$p^J_{\mu}=\sum_i p^i_{\mu}$. 

If the number of the final states particles are large 
the method of moving the trial cone  gives  a slow
algorithm. One can speed it up with starting the cone iteration
at the center of seed towers which passed a minimum cut-off.
But the seeded algorith may have problems
with sensitivity to the emission of soft gluons.
The algorithm allows for jet overlaps, therefore,
one should also specify the  details of
jet merging and jet splitting. This allows for many ad hoc 
options and  it is   difficult to ensure that the definition
of jets are the same at the detector and the parton levels. 
 In a  quantitative analysis  all these  details should be treated with
great care. 

\subsection{$k_T$ algorithms}
The $k_T$ algorithm was designed to avoid jet overlaps and problems
with kinematical boundaries in the case of resummation~\cite{ktalg}.
I summarize the version of $k_T$ algorithm suggested by Ellis and Soper.
The algorithm starts with the initial list of  particles and
an empty list of  jets. Performing the algorithm we get an empty
list of particles and the  list of jets, each separated by $\Delta R^2_{ij}$.
 This regrouping of  particles
into jets goes iteratively through five steps.
\begin{itemize}
\item[i)]
 Every particle (pseudoparticle) $i$ and every particle pair
$(i,j)$  in the particle list
is associated with a $d-$value 
\be
d_i=p^2_{i,T}
 \quad\quad
  d_{ij}\,=\, {\rm min}(p^2_{i,T}\,,\,p^2_{j,T})
\frac{\Delta R^2_{ij}}{D^2}
\ee
where $D$ is a free parameter (the usual choice for its value is $D=1$) and
$\Delta R^2_{ij}$ is the square of the distance between the particles in the
lego-plot
 $\Delta R^2_{ij}=(\eta_i-\eta_j)^2+(\phi_i-\phi_j)^2$.
\item[ii)] 
Find $d_{\rm min}={\rm min}(d_i,d_{ij})$. 
\item[iii)] 
If $d_{\min}=d_{ij}$ replace the particle 
pair $(i,j)$ with a pseudoparticle. Its   momentum $p_{ij}$ is calculated
via the rules of the  recombination scheme. 
For example using the $E$-scheme $p_{ij}\,=\, p_i+p_j$.    
\item[iv)] 
If $d_{\min}=d_i$
remove particle (pseudoparticle) $i$ from the list of particles and
add it to the list of jets. 
\item[v)]
If at this step the particle (pseudoparticle) list ist not empty
got to step i).
\end{itemize}
In the experimental analysis
the $k_T$ algorithm has to combined with some preclustering
procedure in order to keep the jet analysis computionally feasible
and  to diminish the detector dependence of the alghorithm.
The $k_T$ algorithm has the tendency to reconstruct
 more energy from calorimeter
noise, pile-up and underlying event and multi $p\bar{p}$ interactions
than a cone algorithm~\cite{betterjetalgo}. The jet momentum resolution
appears, however,  to be the same for $k_T$ jets and cone jets. Therefore,
it appears that the use of $k_T$ jets has clear overall advantages.

\section{NLO cross sections}
\subsection{Formalism}
The lowest order cross-sections strongly depend on the unphysical
renormalization and factorization scales. Higher order cross sections
reduce this sensitivity  with
a factor of ${\cal O}(\as)$. The calculation of the  NLO corrections
is  technically involved since 
 the virtual and gluon bremsstrahlung 
contributions
 are separately divergent ( soft and collinear
singularities).
 Fortunately, the divergent pieces are universal in the sense that they
 are given by the Born cross sections times a universal
process independent singular factor~\cite{EKS,GGK}. 
The singular parts of the virtual
cross sections can be cancelled analytically with contributions of 
of the real contributions with the use of  local
 counter terms. They  are used to subtract the real contributions
in the singular soft and  collinear regions. 
This subtraction procedure is consistent with numerical 
 Monte Carlo evaluation of the phase space integrals provided one calculates
physical quantities which are defined in terms of infrared
safe measurement functions specified below.
\begin{figure}[htb]
\begin{minipage}[t]{80mm}
\vspace{0cm}
\psfig{figure=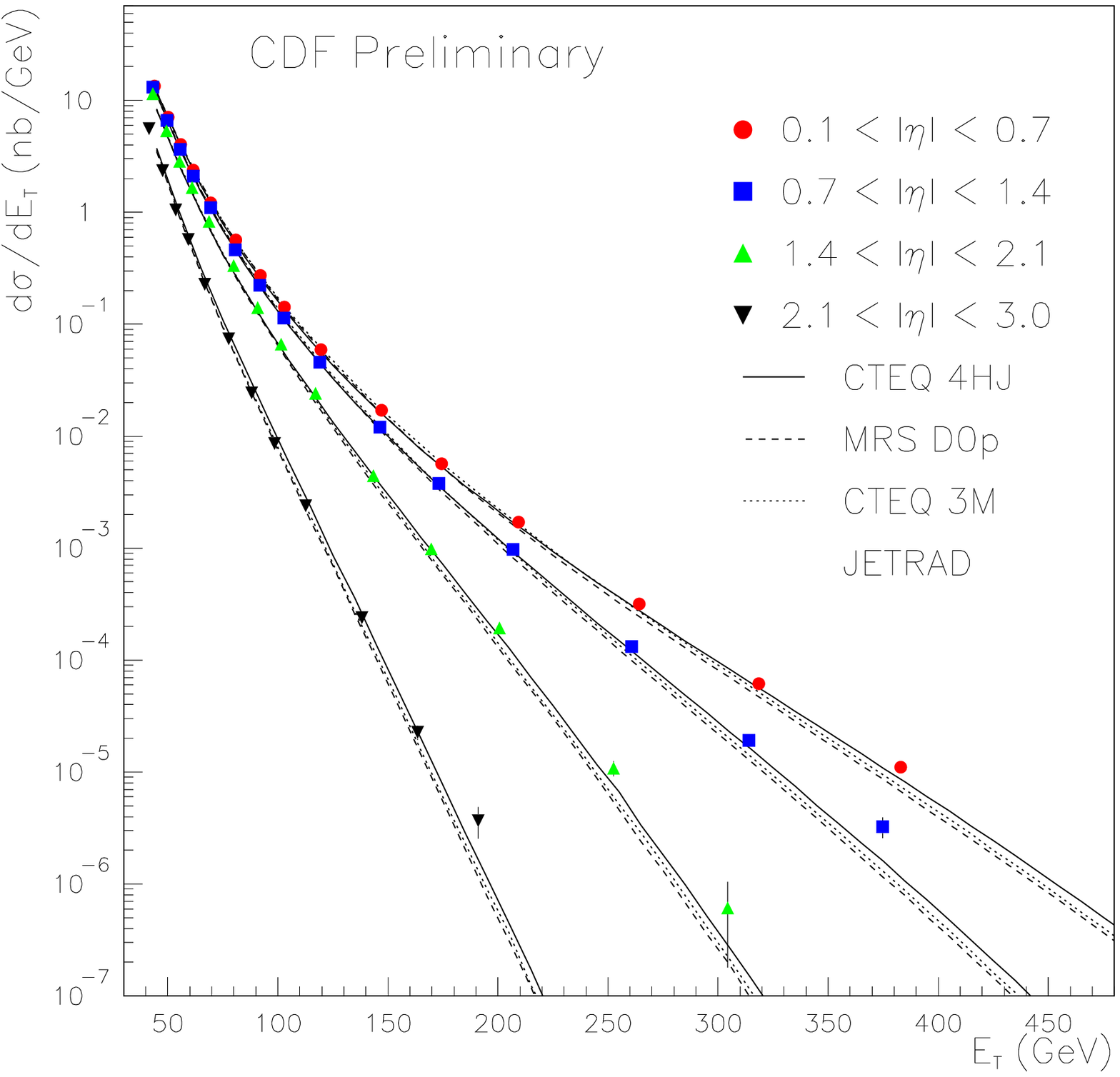,width=7.5cm}
\caption{Inclusive jet cross-section~\cite{Abe:1996wy}
as function of the jet energy
at various pseudo rapidity intervals. The CDF data are
compared with the predictions  calculated by the JETRAD 
NLO MC program~\cite{GGK}.
}
\label{fig:fig1}
\end{minipage}
\hspace{\fill}
\begin{minipage}[t]{60mm}
\vspace{0.8cm}
\epsfig{bbllx=100,bblly=590,bburx=490,bbury=640,
file=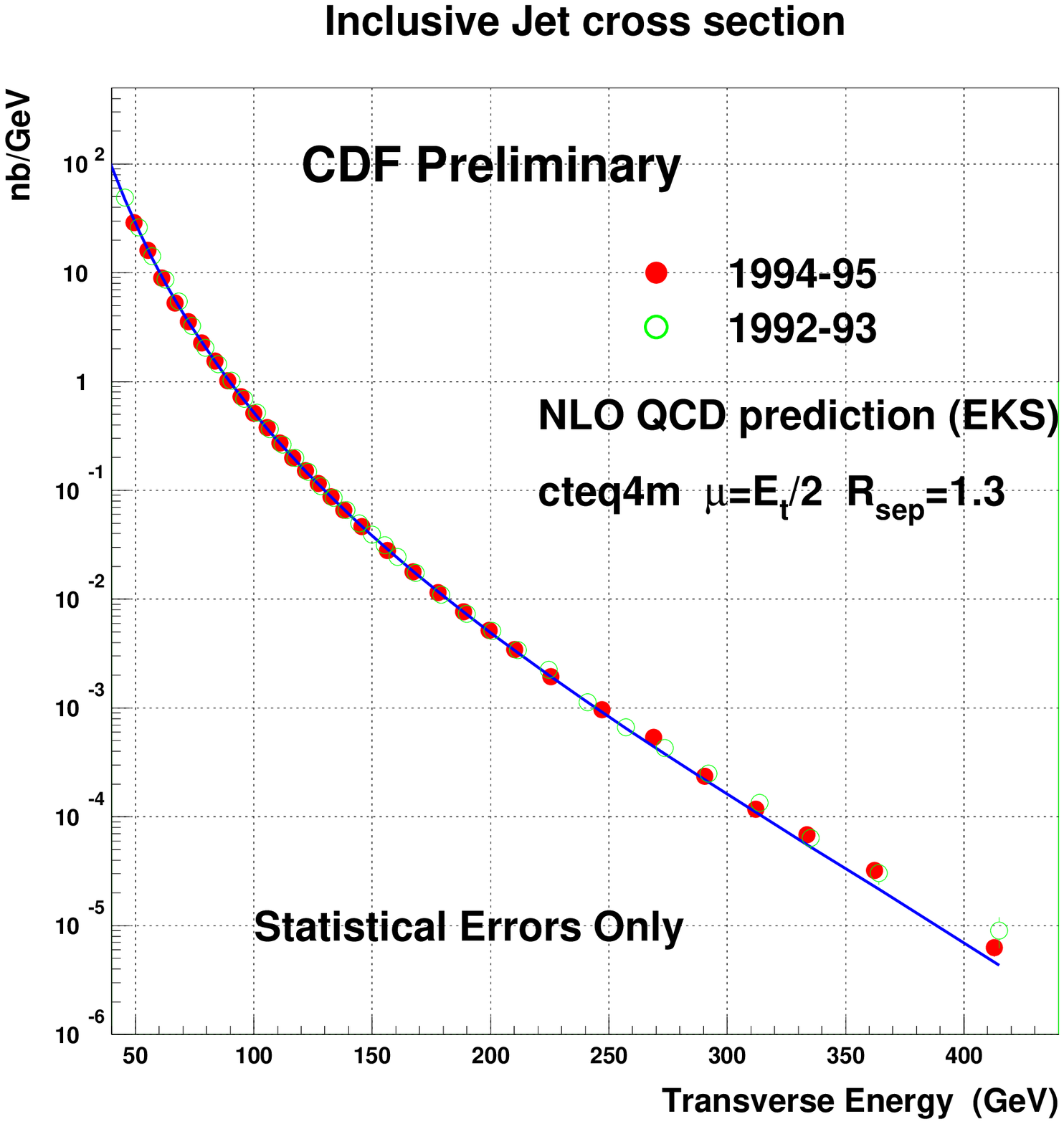,width=55mm}
\vspace{5.5cm}
\caption{Inclusive jet cross-section as function of the
 transverse energy measured by the CDF Collaboration~\cite{Abe:1996wy}. 
The solid curve is the NLO QCD prediction~\cite{EKS}.
}
\label{fig:fig2}
\end{minipage}
\end{figure}
In the case of
 the collisions of hadron A with hadron B, 
the physical cross section is given in terms of parton number
densities $f_{h/H}(x,\mu)$ and parton-parton scattering cross sections
$\hat{\sigma}_{a,b}$  as follows 
\bea
&\sigma(p_A,p_B) = \sum_{a,b} \int dx_a dx_b 
 f_{a/A}(x_b, \mu_F^2) f_{b/B}(x_b, \mu_F^2)\\ \nonumber
&
 \qquad\qquad\qquad\qquad\times 
\left\{\hat{\sigma}_{a,b}^{LO}(x_a p_A, x_b p_B)  +  \hat{\sigma}_{a,b}^{NLO}(x_a p_A, x_b p_B)\right\}
\eea
where for example the   LO cross section for the physical
quantity ${\cal S}$ is given by the phase space integral
over the squared matrix element of the 2-to-n process,  weighted by the 
appropriate measurement function ${\cal S}(p_a,p_b;p_1,...p_n)$
\bea\nonumber
\hat{\sigma}_{a,b}^{LO}( p_a, p_b;[{\cal S}^(n)]) =\int_{n} 
d\hat{\sigma}_a^{B}(p_a,p_b)
=\int  d\Gamma^{(n)}|M_{a,b}|^2{{\cal S}^{(n)}(p_a,p_b,p_1,..,p_n)}
\eea
The finite parton cross sections in NLO are obatained by summing  the
virtual, real and counter term contributions 
\begin{eqnarray}\nonumber
\hat{\sigma}_a^{NLO}(p_a,p_b) = 
\int_{n+1} \hat{\sigma}_{a,b}^{R}(p_a,p_b)+
\int_{n} \hat{\sigma}_{a,b}^{V}(p_a,p_b)+
\int_{n} \hat{\sigma}_{a,b}^{\rm col}(p_a,p_b)
\end{eqnarray}
The individual contributions are  evaluated in 
d-dimension.
In the numerical evaluation of the phase space integral over the real part  
we need to subtract the singular contributions locally.
Fortunately there are several general methods for constructing such local
subtarction terms analytically~\cite{EKS,GGK,CS}.
We can write
\be\nonumber
\label{local}
d\sigma_{a,b}^{NLO} = \left\{d\sigma_{a,b}^R -  d\sigma_{a,b}^{\rm loc}\right\}
+  d\sigma_{a,b}^{\rm loc}  + d\sigma_{a,b}^V + d\sigma_{a,b}^{\rm col}\;\;,
\ee
The local subtraction term $ d\sigma_{a,b}^{\rm loc}$
is a suitable  approximation of $ d\sigma_{a,b}^{R}$. They become equal in the
singular regions. In order to cancel analytically the singular parts
of the virtual contributions we must be able to carry out analytically the
integrals over the  single parton subspaces in $d=2-2\epsilon$ dimension.
After performing this  integration over the local subtraction terms   
the singular pieces of the terms  in eq.~(\ref{local}) 
 can be  cancelled analytically and the
remaning part can be evaluated numerically in four dimensions.
\begin{figure}[htb]
\begin{minipage}[t]{85mm}
\vspace{0.5cm}
\epsfig{file=tfig.3.epsi, width=80mm}
\caption{Comparison of the data and the theory in the case of 
inclusive jet production. Jets are defined by the 
the $k_T$ algorithm~\cite{Abe:1996wy}.  
}
\label{fig:fig3}
\end{minipage}
\hspace{\fill}
\begin{minipage}[t]{70mm}
\vspace{-3cm}
\epsfig{bbllx=100,bblly=390,bburx=490,bbury=640,
file=tfig.4.epsi, width=65mm}
\vspace{5.5cm}
\caption{Theoretical prediction
for inclusive  three jet production at the Tevatron ( Z.~Nagy~\cite{nagy3j}).
The band indicate the theoretical uncertainty due to the variation of the
renormalization and factorization scales $x_{R,F}$ between 0.5 and 2.
The inset shows the K-factor.
}
\label{fig:fig4}
\end{minipage}
\end{figure}
The cancellation mechanism of the soft and collinear singularities
of the virtual corrections against the singular part of the
real contribution
is independent of  the
form of the measurement functions
provided that they 
are insensitive to collinear splitting and soft emission.
That~is in the soft or/and collinear configurations
the measurement functions must fulfil the condition
\be
{\cal S}^{(n+1)}(p_a,..p_{n+1})={\cal S}^{(n)}(\tilde{p}_a,...\tilde{p}_n)
\ee
The  existence 
of universal local subtraction terms~\cite{EKS,GGK,CS} is  crucial for the
method.
The most widely used implementation is the one  by Catani and Seymour\cite{CS}.
The construction of the  NLO parton level Monte Carlo programs 
for 4-leg processes is by now a routine (although rather laborious) work.
The success of the NLO description of jet production  at hadron
colliders is well illustrated by the plots shown figures~1 and 2.
We  see a spectacular agreement in a wide range of
kinematical variables. The  ambiguities due to errors  
in the  parton number densities are shown in figure~3.

\begin{figure}[htb]
\begin{minipage}[t]{75mm}
\vspace{1.cm}
\epsfig{file=tfig.5.epsi, width=75mm}
\vspace{-1.05 cm}
\caption{Comparison of the  scale dependence of the three jet
cross sections obtained in LO and NLO accuracy~\cite{nagy3j}.}
\label{fig:fig5}
\end{minipage}
\hspace{\fill}
\begin{minipage}[t]{70mm}
\vspace{-2.75cm}
\epsfig{bbllx=100,bblly=190,bburx=490,bbury=640,
file=tfig.6.epsi, width=67mm}
\vspace{1.8cm}
\caption{
The NLO prediction for  differential (3+1)-jet cross-section
in the range  $5<Q^2< 5000\gev^2$ is compared
with the data of the H1 collaboration~\cite{nagytrocs3jep}.
}
\label{fig:fig6}
\end{minipage}
\end{figure}

\section{NLO description of 5-leg process}
%%%%%
\subsection{New results for \epem}
The final experimental analysis of 4-jet production
and its comparison with the NLO theoretical predictions was recently
completed~\cite{dissertori}.
 The most remarkable result 
is that we obtain a more precise measurement of the strong coupling
constant from the analysis of the 4-jet data than  from the 3-jet data.
This is not completely suprising since the 
  cross section of four jet production is more sensitive to
 $\alpha_s$ than the three jet cross section, furthemore due to the 
large number of events the statistical accuracy is not the dominating the
experimental error.

\subsection{New results for $pp(\bar{p})$ and $ep$ scattering}
Important  progress  in this topics is  that  the theoretical 
results~\cite{BDK4j} (where the singularities are not manifestly cancelled) 
could be implemented in a efficient parton level C++ Monte Carlo program
(called NLO++ )  
both for $ep$~\cite{nagytrocs3jep} and for $pp(\bar{p})$~\cite{nagy3j}
 scattering.
Similarly to the case of $\epem$ annihilation, the accuracy of the 
theoretical prediction is greatly improved. This   
is illustrated in figure~4 and 5. In fig.~4  inclusive
jet cross section for three jet production is plotted using the $k_T$
algorithm. As we mentioned above, in multi jet production 
the use of the cone algorithm is cumbersome.
It is assumed that the jets are produced in the pseudo rapidity interval
$\abs{\eta}< 4$ and   at  energy of $\sqrt{s}=1800\gev$, with minimum
 transverse jet energy $E_T > 50\gev$. The cross section is plotted
as a function of the transverse energy of the leading jet $E_T^{(1)}$.
The hard scattering factorization and renormalization scale is chosen
to be $E_T^{(1)}$. The theoretical ambiguity is indicated
as a band when the hard scattering scale is changed by a factor in the range 
$0.5<x_{R,F}<2$. One can see that  at NLO the cross section  values 
show much less  sensitivity
to the value of the hard scattering scale as the LO cross sections.
 Hopefully, these predictions can be compared with the Tevatron
data in the near future. In fig.~5 
integrated  cross section values  of 
 three jet production (with the same kinematics as in fig.~4 but
 integrated over $E_T^{(1)}> 100\gev$)  are plotted as a function of the
ratio $\mu/E_T^{(1)}$ where $\mu$ is the value of the hard scattering
scale.

Due to crossing symmetry the $\epem$ NLO four jet matrix elements 
 allow also the
evaluation of the DIS 3+1 jet processes at NLO. The implementation
of the matrix elements~\cite{BDK4j} 
into C++ Monte Carlo program with the corresponding
local subtraction terms  has been 
 recently carried out~\cite{nagytrocs3jep}. The new
result is crucial in  the quantitative comparison of the
data with the theory. As one  can see in fig.~6,   
at NLO,  we get  substantial reduction in the precision of the
theoretical prediction and at the same time the large NLO
corrections are needed to bring the data in agreement with the theory.

\section{NNLO calculations}
In the last few years, 
 the most spectacular  progress  
 has been achieved in developing the technics of calculating
 cross section values  at NNLO accuracy. This requires
 the analytic computation of  two loop corrections to 4-leg QCD processes
(for a  recent more complete review see~\cite{gehrmann-radcor}.)
 The basic ingredient behind this
success is the use of two technical tricks. First,  if one uses 
 integration by parts~\cite{chet} and 
Lorenz invariance~\cite{gr} identities one can 
reduce the very large number of different integrals 
 into a few
master integrals. This step can be very well
computerized and by now  several computer algorithms are available which
perform  automatically this  reduction. Second, all the remaining
master integrals (double box diagrams, etc.) 
could be evaluated analytically. Three different methods have been
used:
 i)~Mellin-Barnes integral transformations~\cite{smirnov}; ii)~
inhomogeneous linear differential equations~\cite{gr} and 
iii)~expansions in terms of nested harmonic sums~\cite{moch}. 
In addition,  
the integrals could be calculated  numerically using 
sector decomposition~\cite{gudrun}. Very recently, 
the differential equation method could be
extended also to phase space integrals~\cite{anmelnik}. 
New classes of functions  appear in the
final solutions. Numerical algorithms are available for the fast evaluation
of these functions.
All parton-parton scattering amplitudes could be calculated analytically
at two loop order.  Analytic expressions of the virtual corrections 
to the cross sections of three jet production
in $\epem$ annihilation have been published recently~\cite{GGGKR}. 
As a first phenomenological application Bern et.al.
\cite{BDS2f2loop}
have computed the cross section of 
  two photon production at LHC in NNLO accuracy. In the case of this process
 the treatment of the bremsstrahlung corrections are considerable
easier than in the general case.
For processes involving
initial hadrons, 
the parton evolution  has  to be evaluated also  at NNLO accuracy.
The results for the non-singlet case have been 
published recently  by
Moch and Vermaseren. This is an outstanding technical achievement. 
If the  NNLO splitting functions~\cite{josnew} will be known
the fitting of  the initial parton density functions can be carried out 
at NNLO accuracy. Depending on 
the nature of the Beyond the Standard Model 
Physics,  the newly achieved theoretical results   may be 
crucial for the
success of the experimental programs at future collider experiments.
The impact of these developments can not be overestimated for the
future high precision measurements at HERA, Tevatron, LHC and TESLA.

\end{document}